\documentstyle[aps,multicol,epsfig]{revtex}  
  
\begin{document}
\title{On statistically stationary homogeneous shear turbulence}
\author{J\"org~Schumacher and Bruno~Eckhardt}
\address{Fachbereich Physik, Philipps-Universit\"at Marburg, D-35032 Marburg, Germany}
\date{\today}
\maketitle

\begin{abstract}
A statistically stationary turbulence with a mean shear gradient is
realized in a flow driven by suitable body forces. The flow domain is
periodic in downstream and spanwise directions and bounded by
stress free surfaces in the normal direction. Except for small
layers near the surfaces the flow is homogeneous. The fluctuations
in turbulent energy are less violent than in the simulations using
remeshing, but the anisotropy on small scales as measured
by the skewness of derivatives is similar and decays weakly with
increasing Reynolds number.
\end{abstract}

\vspace{1cm}
\begin{multicols}{2}
\noindent
{\em Introduction}.---
Although most flows in nature and laboratory are anisotropic on
large scales the statistical behaviour on small scales is expected
to become isotropic \cite{K41,Lum67}. This seems to be supported 
by experiment and
numerical analysis on the level of second order moments 
\cite{Sadd94,Fern95,She93}. 
However,
motivated by analogous behaviour in the passive scalar problem,
Pumir and Shraiman suggested that higher order
moments might remain anisotropic
even for large Reynolds number \cite{Pum95,Pum96}. It would seem 
that a natural situation in which to investigate this problem is that of a 
homogeneous shear flow, in which the time averages $\langle\cdot\rangle$
of the velocity
field are independent of the position in space and satisfy
\begin{equation}
\langle u_x\rangle=S y,\,\,\,\langle u_y\rangle=\langle u_z\rangle=0,
\end{equation}
with (constant) shear rate $S$.
Despite its simple appearance, both experimental and 
numerical realizations of this flow are problematic. 
In the experiment the mean shear is produced by suitable constraints
behind a grid, so that the turbulence is advected and decaying and
eventually influenced by rigid walls. The incompatibility of 
stationarity and homegeneity in such flows has been discussed in detail by
Corrsin {\it et al.} \cite{CHC70,Harr77,Tav81}. 
In numerical simulations using the pseudospectral technique
the mean shear was always 
implemented by the Rogallo remeshing procedure, which corresponds
to a time-periodic driving of the flow \cite{Rog81,Pum96}. Our aim here is to 
present an alternative numerical approach to simulations of
homogeneous shear flows that avoids the periodic driving and 
allows to maintain a statistically stationary state for long times.

{\em Numerical Implementation}.--- 
With lengths measured in units of the gap width $d$, and times
in units of $S^{-1}$, the dimensionless form of the 
equations for an incompressible Navier--Stokes
fluid become  
\begin{eqnarray}
\label{nseq}
\frac{\partial{\bf u}}{\partial t}+({\bf u}\cdot{\bf \nabla}){\bf u}
&=&-{\bf \nabla} p+\frac{1}{Re_s}{\bf \nabla}^2{\bf u}+{\bf f}\;,\\
{\bf \nabla}\cdot{\bf u}&=&0\;
\end{eqnarray}
where $p({\bf x},t)$ is the pressure,
${\bf u}({\bf x},t)$ the velocity field. The shear 
Reynolds number is
$Re_s=S\,d^2/\nu$ with $\nu$ as kinematic viscosity. 
In the $x$ (streamwise) and $z$ (spanwise) directions periodic boundary
conditions apply. In the other direction the    
 flow domain is bounded by two parallel flat surfaces
that are assumed to be
impenetrable and stress-free, 
{\it i.e.}  $u_y=\partial_y u_x=\partial_y u_z=0$.  
The effects of the free slip surfaces at $y=0$ and $y=d$ on the bulk
behaviour are much weaker than those of rigid walls, since only
the wall-normal component is forced to vanish. The resulting boundary
layer in the wall normal component is of the order 
$\nu/u'_{y,\,rms}$ where 
$u'_{y,\,rms}=\langle(u'_y)^2\rangle^{1/2}$ denotes
the root mean square fluctuations of the turbulent wall normal velocity.
The statistical properties of the tangential
components are not affected by this boundary layer.

The mean shear and turbulence are maintained by a suitable
body force ${\bf f}({\bf x},t)$. A linear mean profile 
$\langle u_x\rangle(y)=(1/2-y)$ for $y\in [0,1]$ 
 can be approximated by a finite Fourier sum of cosines
\begin{eqnarray}
\langle u_x\rangle(y)\simeq\frac{4}{\pi^2}\sum^{5}_{n=0} 
\frac{\cos[(2n+1)\pi y]}{(2n+1)^2}\,.
\label{profile}
\end{eqnarray}
The external forcing ${\bf f}$ was chosen such that the six modes used in 
eq.~(\ref{profile}) remained constant in time, {\it i.e.} 
$\partial_t \mbox{Re\,} [u_x({\bf q},t)]=0$ for Fourier modes with 
${\bf q}=(2n+1)\pi {\bf e}_y$ for $n=0$ to 5. At sufficiently high shear
rates the flow is unstable and turbulence sets in. Then the force
${\bf f}$ fluctuates in time as well.

The free slip boundary conditions allow for efficient numerical
simulations with Fourier modes for the velocity components.
Nevertheless, the data presented here amount to approximately 360 CPU hours of
computing time on a Cray T-90.
The equations are integrated by means of a
pseudospectral technique using a 2/3-rule dealiazing \cite{Canuto}.
The integration domain has an aspect
ratio $L_x:L_y:L_z=2\pi d:d:2\pi d$ and is resolved 
by $256\times 65\times 256$ Fourier modes.  

\begin{figure}
\begin{center}
\epsfig{file=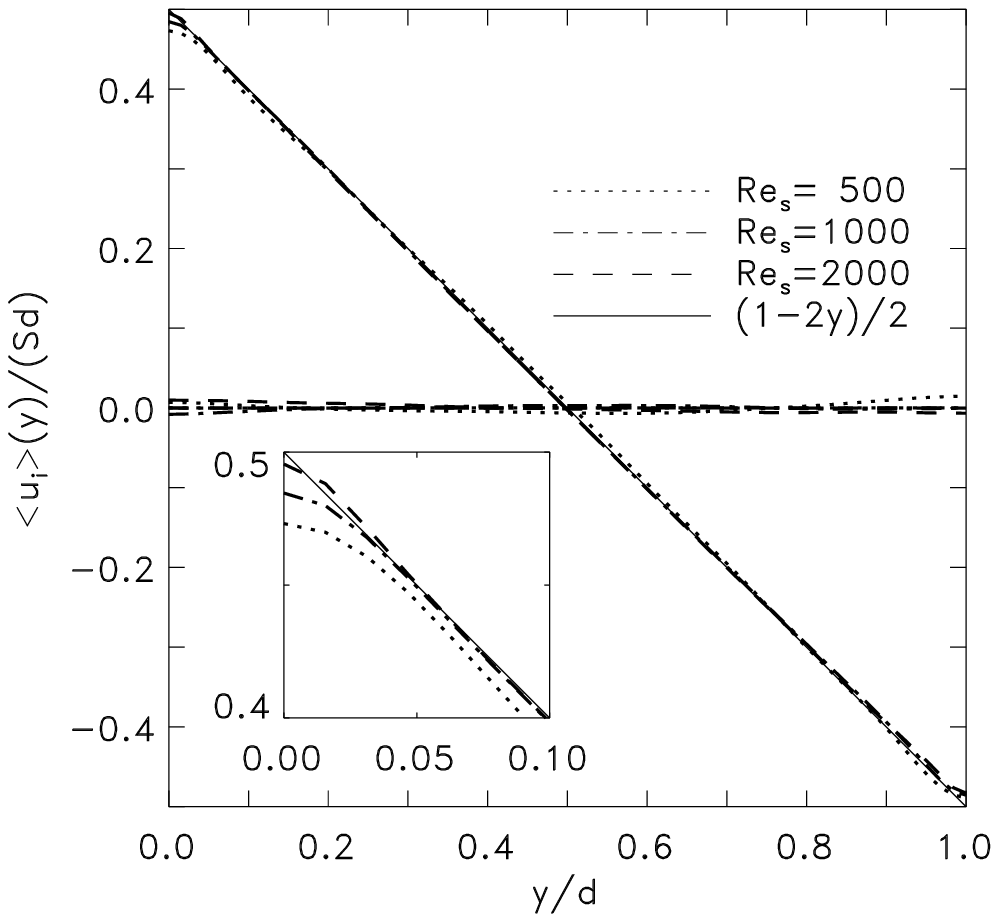,width=6.5cm,height=7.3cm}
\hspace{0.5cm}
\epsfig{file=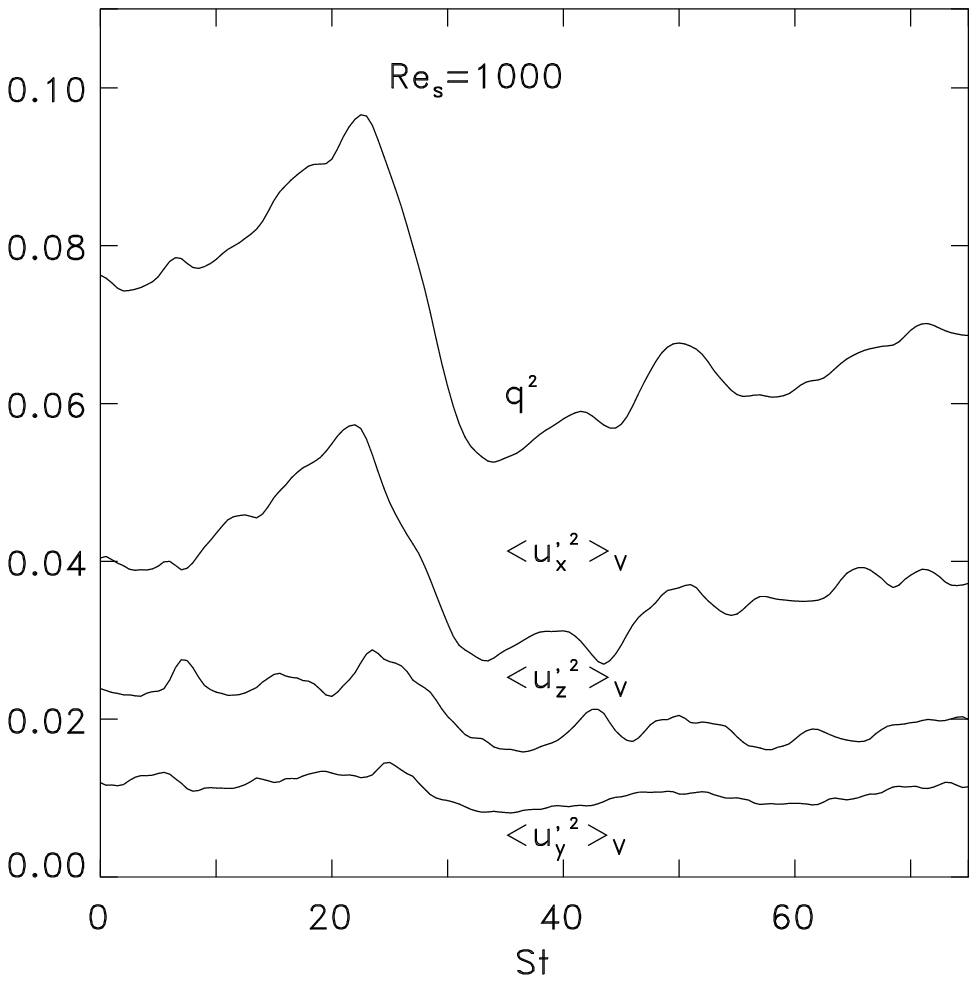,width=6.5cm,height=7cm}
\end{center}
\vspace{0.5cm}
\caption[]{The mean velocity profiles $\langle u_i\rangle(y)$ for the
three components $i=x, y$ and $z$,
averaged in time and over planes normal to the mean shear, 
for $Re_s=500,\,1000$ and $2000$. 
The solid line is the idealized linear 
function in dimensionless form. The inset resolves the small layer
near the surface where the mean shear is not constant.}
\label{p1}

\vspace{0.5cm}
\caption[]{Time traces of the turbulent kinetic energy 
$q^2(t)/(Sd)^2$ and of the separate contributions 
from the turbulent velocity components $\langle (u'_i)^{2}\rangle_V/(Sd)^2$
for the three components $i=x, y$ and $z$. }
\label{p2}

\vspace{0.5cm}
\caption[]{Isosurfaces of the streamwise turbulent velocity component $u'_x$
for $Re_s=500$ at the levels $u'_x/(Sd)=\pm 0.48$. The positive value 
is brighter.}
\label{p3}

\vspace{0.5cm}
\caption[]{Same as fig.~\ref{p3}, but for $Re_s=2000$ 
with levels $u'_x/(Sd)=\pm 0.46$. At this higher
Reynolds number the flow structures are more fragmented
and smaller. }
\label{p4}
\end{figure}

{\em Stationarity}.---  
When averaged in downstream and spanwise direction
as well as in time, the mean velocity components show the
expected shear flow behaviour, $\langle u_y\rangle = \langle u_z\rangle=0$
and $\langle u_x\rangle\sim - y$ (see fig.~\ref{p1}). 
The downstream profile
differs from the linear shear flow only in a small region
near the surfaces; this region decreases as the 
Reynolds number increases (inset of fig.~\ref{p1}). Before starting the 
statistical analysis a forward integration over a 
period of $ST\ge 20$ (time is measured in units of the shear rate $S$)
with the full spectral resolution
was always performed to guarantee relaxation to the turbulent state.

The fluctuations in the velocity field are defined as
$u'_i=u_i-\langle u_i\rangle$ for $i=x, y, z$. Figure~\ref{p2}
shows the time evolution of the total kinetic energy in the 
fluctuations, 
$q^2(t)=\langle (u'_i)^2\rangle_V$, where $\langle\cdot\rangle_V$ denotes
an average over the volume. The amplitude of the variations in
kinetic energy decreases with increasing Reynolds number. This seems
to be connected to the fragmentation of coherent streaks and vortices 
\cite{Schu00,Wal97}
with increasing $Re$ and the reduced downstream correlation. 
This is shown by volume surface plots of the turbulent streamwise
velocity component at the lowest (see fig.~\ref{p3}) and the
highest (see fig.~\ref{p4}) of our Reynolds numbers.
The periods of the oscillations in total kinetic energy 
are surprisingly large, thus requiring very long time integrations
for converged time averages. Compared to the long time simulations of
Pumir \cite{Pum96} the fluctuations are smaller in amplitude
and do not show the violent bursts.

Finally, statistical stationarity implies balancing of the
turbulent kinetic energy in the mean,
\begin{eqnarray}
0=
-\nu\langle[\partial_i u'_j(\partial_i u'_j+\partial_j u'_i)]\rangle
-\langle u'_x u'_y\rangle \partial_y\langle u_x\rangle
+\langle u'_x f'_x\rangle.
\end{eqnarray}
The first term on the right side is the energy dissipation rate $\epsilon$
and the second term is the turbulent energy production rate $P$. The last term,
is the energy injection due to the applied volume forcing. 
In the bulk, outside the
small boundary layers near the top and bottom surfaces,  we find 
that the production and dissipation differ by less than 4$\%$ and that 
the energy injection from the volume forcing is negligibly small. 
Here we took volume and time averages for the Reynolds stress component
in the production term, for the energy dissipation rate and the
energy injection term. We used an averaging time in shear rate 
units, {\it i.e.} $ST$,
of 75 for the lower $R_{\lambda}$ and 50 as well as 100
for the highest $R_{\lambda}$. 

{\em Higher order statistical moments}.--- 
Data on statistical moments from the experiment of 
Garg and Warhaft \cite{Garg98}, two sets of direct numerical simulations
\cite{Pum96} and \cite{Moin87} and our simulations are collected in table~I.
The data sets cover almost the same range of parameters with similar
behaviour, despite the different realizations of the mean shear.
The dimensionless $S^{\ast}=S q^2/\epsilon$ does not seem to
vary with $Re_s$, but $S(\nu/\epsilon)^{1/2}$ decreases for our range
of $Re_s$. This agrees with the observations of Pumir and Shraiman
\cite{Pum95,Pum96}.
\end{multicols}
\begin{table}
\renewcommand{\arraystretch}{1.1}
\begin{tabular}{lcccccc}
 & Exp.~\cite{Garg98} &  DNS~\cite{Moin87} & DNS~\cite{Pum96} & $Re_s=500$ & $Re_s=1000$ & $Re_s=2000$\\
\hline
$\langle (u'_x)^{2}\rangle/q^2$                  &  -- & 0.53& 0.53& 0.57& 0.55& 0.52 (0.52)  \\ 
$\langle (u'_y)^{2}\rangle/q^2$                  &  -- & 0.16& 0.21& 0.12& 0.15& 0.18 (0.17)  \\ 
$\langle (u'_z)^{2}\rangle/q^2$                  &  -- & 0.31& 0.26& 0.31& 0.30& 0.30 (0.31)  \\ 
$S_1^{\ast}=S\langle (u'_x)^{2}\rangle/\epsilon$ & 3.31& 4.58& 3.98& 4.17 &  4.72& 4.20 (4.26)\\
$S^{\ast}=S q^2/\epsilon$   &  -- & 8.65& 7.50& 7.32 &  8.58& 8.08 (8.19)\\
$S(\nu/\epsilon)^{1/2}$                          &0.04 & --  & 0.11& 0.38 &  0.25&0.18 (0.17)\\
$P/\epsilon$                                     &0.96 & --  & --  & 1.04 & 0.99 &1.02 (0.98)\\
$R_{\lambda}$                                    & 310 &  73 & 90  & 59 &  87 & 94 (106)     \\ 
$S_{\omega_z}$                                   & --  &  -- & -0.58&-0.79 & -0.67 & -0.57 (-0.60) \\
$K_{\omega_z}$                                   & --  &  -- & --   & 5.62 & 6.57 & 7.36 (7.45) \\
$S_{\partial u'_x/\partial y}$                   & 0.5 &  -- & 0.87 & 0.96 & 0.90 & 0.82 (0.84) \\
$K_{\partial u'_x/\partial y}$                   & 8.6 &  -- & --   & 5.59 & 6.38 & 7.08 (7.14)\\
\end{tabular}
\caption{Comparison between experiment and different simulations on homogeneous 
shear turbulence. 
The Taylor-Reynolds number is calculated from
$R_{\lambda}=\langle (u'_x)^{2}\rangle/\nu
[\langle(\partial_x u'_x)^2\rangle]^{1/2}$\,.
The skewness and kurtosis of a field $\phi$ are defined as 
$S_{\phi}=\langle\phi^3\rangle/\langle\phi^2\rangle^{3/2}$ and 
$K_{\phi}=\langle\phi^4\rangle/\langle\phi^2\rangle^2$, respectively.
Note that here the total kinetic energy $q^2=\langle (u'_i)^2\rangle$ is
also averaged in time. The values  given in
parentheses in the last column are obtained by averaging over a time interval
twice as long.}
\end{table}

\begin{multicols}{2}
Most quantities in table I change monotonically with $Re_s$. We attribute
the few exceptions to pecularities of the simulations at $Re_s=500$ which
has the strongest coherent structures and is perhaps most strongly
affected by the downstream periodicity.
Coherent structures (predominantly
streaks and downstream vortices)
show up in all our simulations, even at the lowest Reynolds number.
In the simulations of \cite{Lee90} that used remeshing
they were observed for higher values of $S^{\ast}$ only where
the flow became strongly non-stationary.   
The effects of vortices and downstream periodicity may also explain the
rather large skewness values $S_{\omega_z}$ and
$S_{\partial u'_x/\partial y}$ for $Re_s=500$. Both quantities decrease
steadily with $R_{\lambda}$, while they remained constant
($\sim R_{\lambda}^0$) in the simulations of Ref. \cite{Pum96}. The shear 
flow experiments \cite{Garg98} have been fitted to $\sim R_{\lambda}^{-0.6}$
for $150\lesssim R_{\lambda} \lesssim 400$. The statistical error of the data
may be gauged by comparing the results for two different averaging times.
The last two columns in table I, obtained by averaging over $ST=50$ and 
$ST=100$, respectively, show small variation. In particular, 
the trend observed for skewness
and kurtosis is weak but lies outside these statistical variations.  
A similarly slow decay of anisotropy 
effects was also found recently in a systematic analysis of the 
anisotropic scaling contributions
to high-order structure functions in
a turbulent atmospheric boundary layer with $R_{\lambda}\sim 2000$ 
\cite{Sreeni00}.

\begin{figure}
\begin{center}
\epsfig{file=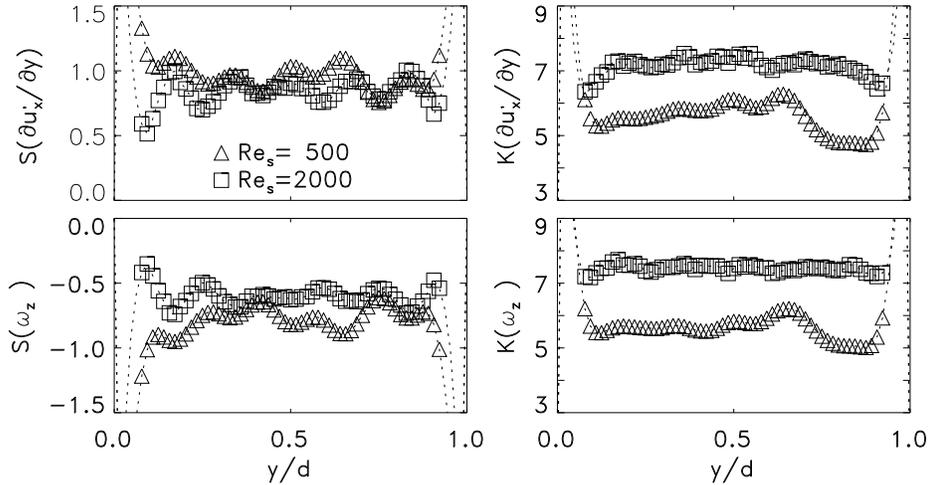,width=12cm}
\end{center}
\vspace{1cm}
\caption[]{Profiles of skewness and kurtosis in normal direction for two
values of $Re_s$. The datapoints closest to the boundary surfaces 
(indicated by dotted
lines) are excluded from the statistical analysis that leads
to the entry in table.~I.}
\label{p5}
\end{figure}

The effects of the surfaces on the higher order moments are
limited to the boundary layers very close to the surfaces.  
The spatially resolved plots in fig.~\ref{p5} show
the shear direction dependence of the
skewness and kurtosis of the spanwise vorticity, 
$\omega_z=\partial_x u'_y-\partial_y u'_x$, and the shear gradient
$\partial_y u'_x$. 
The variations
across the shear layer become smaller with increasing 
Reynolds number.

{\em Summary} ---
The simulations for the statistically stationary shear flows bounded by 
free slip surfaces show that in the central region an approximately
homogeneous shear flow with statistically stationary properties develops.
The moments of the velocity field are compatible with previous
experimental and numerical findings. The most noticable difference 
to the long time simulations by Pumir \cite{Pum96} is the absence of
violent bursts in turbulent energy. 
Further investigations of the statistical properties
of this model for almost homogeneous shear flows are in progress.

{\em Acknowledgments}.---  We thank L.  Biferale, D.  Lohse, M.  Nelkin, K.
Sreenivasan, and F.  Toschi for fruitful discussions,
A. Pumir for comments on the manuscript, and the Institute for
Theoretical Physics at Santa Barbara for hospitality.  This work was supported
in part by National Science Foundation under Grant No.  PHY94-07194 and the
European Union
within the CARTUM project.  The numerical simulations were done on a 
Cray T-90 at
the John von Neumann-Institut f\"ur Computing at the Forschungszentrum J\"ulich
and we are grateful for their support.

\end{multicols}
\clearpage


\begin{thebibliography}{10}
\bibitem{K41} KOLMOGOROV A. N., {\it Dokl. Akad. Nauk SSSR}, {\bf 30} (1941) 301.

\bibitem{Lum67}  LUMLEY J. L., {\it Phys. Fluids}, {\bf 10} (1967) 855.

\bibitem{Sadd94} SADDOUGHI S. G. and VEERAVALLI S. V., {\it J. Fluid Mech.}, 
                {\bf 268} (1994) 333.
                 
\bibitem{Fern95} FERNHOLZ H. H., KRAUSE E., NOCKERMANN M. and SCHOBER M.,
                 {\it Phys. Fluids}, {\bf 7} (1995) 1275.
                 
\bibitem{She93} SHE Z., CHEN S., DOOLEN G., KRAICHNAN R. H. and ORSZAG S. A.,
                 {\it Phys. Rev. Lett.}, {\bf 70} (1993) 3251.                  
                                   
\bibitem{Pum95} PUMIR A. and SHRAIMAN B. I., {\it Phys. Rev. Lett.} {\bf 75} 
                (1995) 3114.
                                                                  
\bibitem{Pum96} PUMIR A., {\it Phys. Fluids}, {\bf 8} (1996) 3112.
                  

\bibitem{CHC70}   CHAMPAGNE F. H., HARRIS V. G. and CORRSIN S., 
                  {\it J. Fluid Mech.}, {\bf 41} (1970) 81.
                  
\bibitem{Harr77}  HARRIS V. G., GRAHAM J. A. H. and CORRSIN S.,                  
                 {\it  J. Fluid Mech.}, {\bf 81} (1977) 657.
                  
\bibitem{Tav81}   TAVOULARIS S. and CORRSIN S.,                  
                  {\it J. Fluid Mech.}, {\bf 104} (1981) 311.
                  
\bibitem{Rog81}   ROGALLO R. S., {\em Numerical experiments in homogeneous 
                  turbulence}, {\it NASA TM} 81315 (1981).
                  
\bibitem{Canuto}  CANUTO C., HUSSAINI M. Y., QUATERONI A. and ZANG T. A.,
                  {\em Spectral Methods in Fluid Dynamics}, Springer, Berlin (1988). 

\bibitem{Schu00}  SCHUMACHER J. and ECKHARDT B., {\em Evolution of turbulent
                  spots in a plane shear flow}, submitted (2000).

\bibitem{Wal97}   WALEFFE F., {\it Phys. Fluids}, {\bf 9} (1997) 883.

\bibitem{Garg98}  GARG S. and WARHAFT Z., {\it Phys. Fluids}, {\bf 10} 
                  (1998) 662.

\bibitem{Moin87}  ROGERS M. M. and MOIN P., {\it J. Fluid Mech.}, {\bf 176}
                  (1987) 33.
                  
\bibitem{Lee90}  LEE M. J., KIM J. and MOIN P., {\it J. Fluid Mech.}, {\bf 216}
                  (1990) 561.
                  
\bibitem{Sreeni00} KURIEN S. and SREENIVASAN  K. R., {\it Phys. Rev. E}, {\bf 62}
                   (2000) 2206.                  
\end{thebibliography}
\end{document}